# Probing the single neurotransmitters with the WGM microcavity-hybridized plasmonic nanospiked antennas

A. K. Arunkumar [1, *], E. Zossimova,[1, 2], M. Walter[2] , S. Pedireddy[1] , J. Xavier[1, 3, *], F. Vollmer[1, *]

## Abstract

Discerning the neurotransmitter dysregulation is a hallmark of neurological disorders and diseases, including Alzheimer's, Parkinson's, and multiple sclerosis. The concentration of neurotransmitters in the synaptic cleft is particularly low, ranging from nM to fM, which makes it challenging to accurately monitor changes over the course of a clinical trial using existing sensing techniques. By means of an advanced whispering gallery mode (WGM) sensor hybridized with plasmonic nanospiked antennas, we detect and discriminate between different neurotransmitters at the single-molecule level. Our results show that the sensor can detect neurotransmitters with exceptional sensitivity down to 10 aM and discriminate between structurally similar neurotransmitters, such as GABA and glutamate, over a large number of detection events. Furthermore, we find that the average WGM resonance shift, induced by a neurotransmitter binding to the sensor, strongly correlates with molecular polarizability values obtained from electronic structure calculations. These findings establish the optoplasmonic WGM sensors as potential biosensor platform in different avenues of neuroscience by detecting and discriminating neurotransmitters as well as investigating their dynamics at ultra low-level concentrations, plausibly contributing to deeper understanding of brain function and neurological disorders.

## Introduction

Neurotransmitters are fundamental chemical messengers that orchestrate synaptic transmission and underpin a wide range of physiological and cognitive processes, including memory, mood regulation, and autonomic control[1, 2, 3, 4, 5]. Dysregulation in their signalling is implicated in numerous neurological and psychiatric disorders such as Alzheimer's disease [6], Parkinson's disease [2], and depression [7], highlighting the imperative need for sensitive and selective tools for their detection, discrimination and quantification.  Advances in detecting, quantifying and distinguishing these neurotransmitters, especially at single-molecule level in a label free manner is essential for improving our understanding of these complex disorders and developing effective treatments. Neurotransmitters are essential to numerous therapeutic and diagnostic applications, prompting extensive research into methods for their detection. Among the most widely used techniques are nuclear medicine imaging, such as positron emission tomography (PET) [8] and single-photon emission computed tomography (SPECT) [9], alongside optical approaches like surface-enhanced Raman spectroscopy (SERS)[10], fluorescence, Förster resonance energy transfer (FRET) [11], chemiluminescence [12] and optical fibre biosensing [13] . Other promising methods include electrochemical techniques, fast-scan cyclic voltammetry (FSCV), amperometry, high-performance liquid chromatography (HPLC), and mass spectrometry [14] . Despite the advances in these detection strategies, significant challenges remain due to the low concentrations that can vary from a peak concentration of millimolar to least attomolar concentrations [15, 16, 17, 18, 19, 20] and rapid turnover of neurotransmitters occurring within a time span of 0.1-1 ms[20], which complicate accurate measurement. Many traditional techniques, such as FRET[11], rely on molecular labelling, which can alter the natural behaviour of neurotransmitters or introduce additional complexities in the labelling process. Furthermore, methods like HPLC, an ubiquitous technique in microdialysis to quantify neurotransmitter levels, are time-consuming and can suffer from low sensitivity, lengthy sample preparation, and poor sample stability[21]. Although newer techniques, such as capillary electrophoresis [22]and mass spectrometry [23], offer some advantages, they still face limitations related to calibration issues and detection sensitivity.

[1] Department of Physics and Astronomy, Living Systems institute, University of Exeter, Exeter, UK. [2]Freiburg Center for Interactive Materials and Bioinspired Technologies (FIT), University of Freiburg, Freiburg, Germany. [3]Currently with SeNSE, Indian Institute of Technology, Delhi, India. E-mail : a.kakkanattu-arunkumar2@exeter.ac.uk; j.xavier@exeter.ac.uk; f.vollmer@exeter.ac.uk

The need for real-time, highly sensitive detection is particularly crucial for monitoring rapidly decaying neurotransmitters like glutamate, which are present in low concentrations in the brain. While SERS offers a potential label-free solution [24, 25], its application remains challenging due to the very low concentrations of neurotransmitters in biological fluids, and the weak Raman signals these molecules produce. Consequently, there is a need for novel approaches that can provide real-time, ultrasensitive detection with improved limits of detection to advance the study of neurological functions and diseases.

Label-free detection of neurotransmitters at the single-molecule level presents significant challenges due to their small molecular size (< 1 nm), extremely low concentrations in biological systems (nM-fM range), and rapid clearance from synaptic clefts (0.1-1 ms). To address these challenges, we developed a highly sensitive optoplasmonic WGM sensor platform capable of detecting individual neurotransmitter molecules with exceptional temporal resolution.

Here, we present an ultrasensitive optoplasmonic WGM biosensor for single-molecule detection of neurotransmitters across a dynamic concentration range from 1 μM to 10 aM. By integrating morphologically engineered gold nanostars (AuNSs) into the sensor design, we achieve enhanced near-field coupling and localized field confinement, significantly boosting detection sensitivity and signal throughput. Notably, we demonstrate molecular discrimination between neurotransmitters with comparable molecular weights, such as γ-aminobutyric acid (GABA) and glutamate, overcoming a key limitation in current sensing modalities. Our findings establish this hybrid WGM-AuNS sensor as a powerful platform for probing neurochemical dynamics with unprecedented sensitivity and selectivity, opening new avenues for investigating the molecular underpinnings of brain function and dysfunction.

## Experimental Results

The detection performance of the microsphere-based sensing platform was evaluated for three key neurotransmitters: γ-aminobutyric acid (GABA), glutamate and dopamine. To enable nanoscale optical field confinement, CTAB-capped gold nanoparticles, specifically nanorods (AuNRs) and nanostars (AuNSs), were immobilized onto the WGM sensor's surface of a glass microsphere (~ 85-90 μm in diameter) (see supplementary Fig. S2), enhancing local field intensity through localized surface plasmon resonance (LSPR) coupling with the WGM's evanescent field. In addition, the detuned surface resonance peaks of CTAB-AuNRs (767 nm) and CTAB-AuNSs (702 nm) (see Supplementary Fig.S1) enables efficient coupling to WGMs. Experimental setup and functionalization of nanoparticles are detailed in the *Methods* section 1, 2 and 3. Redshifts in WGM resonance were observed upon neurotransmitter binding, attributed to local refractive index changes at LSPR hotspots at nanoparticle tips, with responses appearing as either discrete step-like or transient spike-like spectral shifts depending on interaction kinetics. Comparative measurements revealed distinct detection signatures for each neurotransmitter across AuNR- and AuNS-functionalized sensors. All measurements were performed in phosphate buffer (pH 7.4) to emulate physiological ionic conditions (except dopamine), with independent sensor assemblies employed for each neurotransmitter.

At pH 7.4, phosphate anions ($HPO_4^{2-}$ and $H_2PO_4^-$) adsorb onto the gold nanoparticle surface, imparting a net negative surface charge [26]. This facilitates hydrogen bonding with the protonated amine groups ($-NH_3^+$) of GABA ($pK_a$ = 10.5) and glutamate ($pK_a$ = 9.67), leading to stable adsorption of these neurotransmitters at the sensing hotspots. The result is a step-like shift in the resonance trace, indicative of irreversible single-molecule binding events. As shown in Fig.1b, the binding of GABA and glutamate produces discrete redshifts in the WGM resonance, with each step corresponding to a single-molecule event. This is further corroborated by the time-interval distribution of binding events, which follows a single-exponential decay characteristic of Poisson statistics (see Supplementary Fig. S3). The detection was achieved at concentrations as low as 10 aM, well below the detection limits of current label-free single-molecule platforms.

In contrast, dopamine presents a distinct challenge due to its known tendency to undergo spontaneous polymerization into polydopamine under neutral or basic conditions [27] [28]. To circumvent this, dopamine sensing was performed in PBS at pH 6.5, below its $pK_a$ (8.71), ensuring that dopamine remains in a protonated, non-polymerizing state.

To further enhance the sensitivity of neurotransmitter detection, we functionalized the microsphere-based WGM sensors with gold nanostars (AuNSs) and conducted separate measurements with GABA, glutamate, and dopamine, as shown in Fig.1c. Compared to nanorods, AuNSs present a higher density of sharp protrusions, giving rise to a greater number of electromagnetic hotspots and thereby increasing the probability and intensity of signal-generating binding events. Fig. 1d presents histograms of discrete resonance shifts ("step heights") corresponding to single-molecule binding events for each neurotransmitter. At 10 aM concentration, all three analytes showed a substantially higher number of detectable events with AuNSs compared to AuNRs. Quantitatively, the number of detected events increased by 40.6% for GABA, 34.6% for glutamate, and 51.8% for dopamine. This increase highlights the advantage of AuNSs in providing more plasmonically active binding sites. Furthermore, mean step heights, indicative of the average resonance shift per binding event, were consistently higher for AuNS-based sensors across all neurotransmitters. For GABA, the mean step height increased from 3.1 ± 1.1 fm (AuNRs) to 4.5 ± 2.4 fm (AuNSs). Glutamate showed an increase from 4.6 ± 1.9 fm to 5.2 ± 1.7 fm, and dopamine from 4.8 ± 1.8 fm to 5.7 ± 1.6 fm. These results establish the superior detection efficiency of AuNSs, reinforcing their role as optimal plasmonic enhancers for single-molecule sensing using WGM resonators.

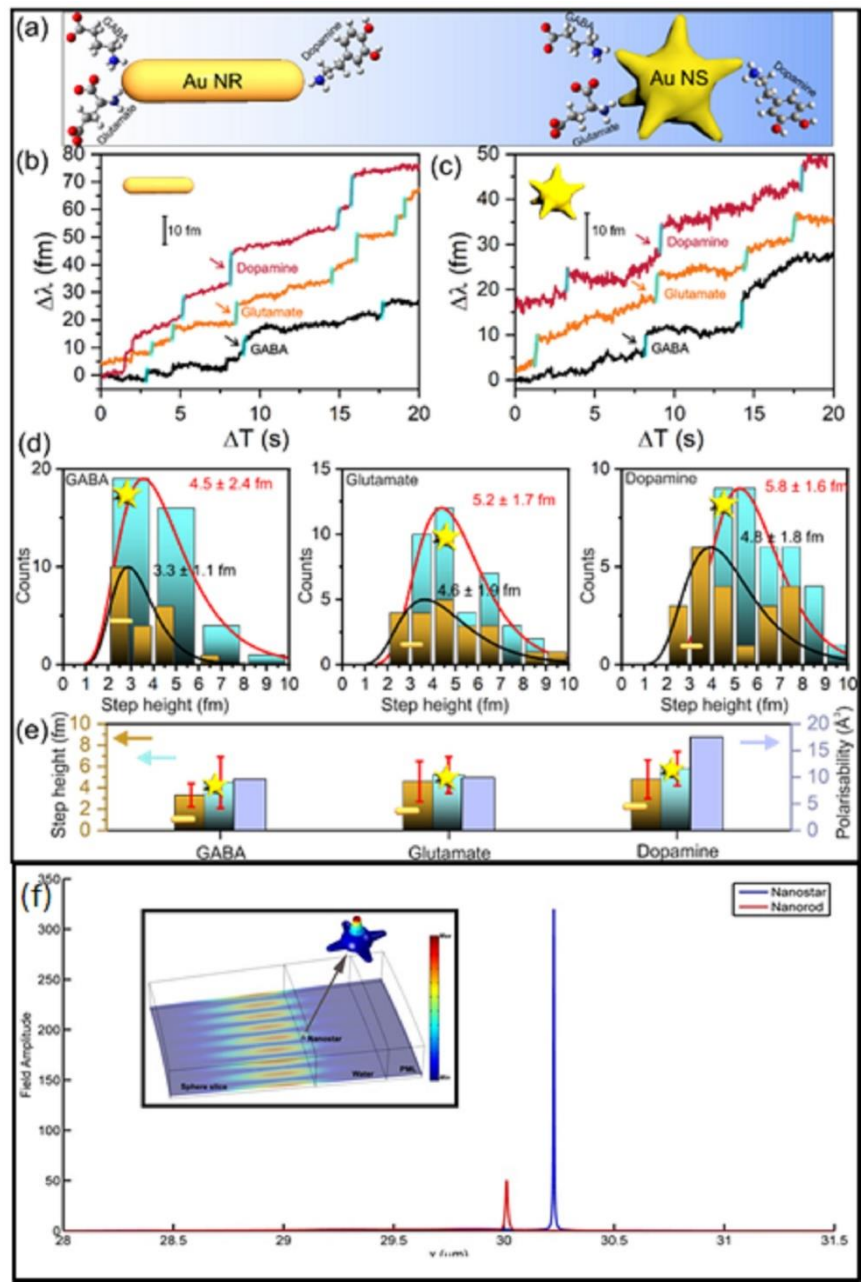

**Fig. 1 | Detection of GABA, glutamate and dopamine using optoplasmonic WGM sensors at 10 aM concentration. a,** Schematic illustration of neurotransmitter GABA, glutamate and dopamine binding with AuNRs and AuNSs respectively. **b,** Time traces of whispering-gallery mode (WGM) resonance shifts (Δλ) showing discrete binding events for GABA (black), glutamate (orange), and dopamine (red) with gold nanorods (AuNRs) **c** Corresponding resonance traces obtained using gold nanostars (AuNSs). Discrete step signals are highlighted in cyan. **d,** Histograms of step heights for GABA, glutamate and dopamine using AuNRs (brown) and AuNSs (cyan), fitted with log-normal distributions. **d,** Histograms of amplitude of step signals for the three neurotransmitters with AuNRs and AuNSs. **e,** Bar chart comparing average step heights for GABA, glutamate, and dopamine on AuNRs and AuNSs, alongside their molecular polarizabilities. **f,** Rigorous FEM analysis of Field distribution of a 100 nm core radius plasmonic nanostar -WGM hybrid microresonator setup with 60 µm diameter silica sphere slice.

The step height distributions were best fit using a log-normal model, consistent with a multiplicative process influenced by multiple underlying factors [29]. The right-skewed nature of the distributions reflects heterogeneity in: (i) Binding site locations-events originating from hotspots (e.g., nanoparticle tips) yield larger signals than those from low-field regions. (ii) Number of active nanoparticles per sensor-typically 3–4 AuNPs are involved, and their collective interaction with the WGM field contributes to signal variability. Other sources of dispersion include the binding orientation of nanorods relative to the WGM polarization, positional heterogeneity on the microsphere surface (e.g., equatorial vs. polar attachment), and geometric asymmetries in the resonator cavity. Such factors introduce a distribution of coupling strengths and electric field enhancements, leading to broadening in the observed step height statistics [30,31] . Despite these variabilities, the average step heights scale with the molecular polarizability of each neurotransmitter, a trend captured in Fig 1e. This correlations supports predictions from recent hybrid quantum classical simulations[32], suggesting that in addition to detecting the presence of single molecules, the optoplasmonic WGM sensor is sensitive to their inherent electronic properties. Fig. 1f shows the FEM simulations - lightning rods of plasmonic nanostars with multiple branched spikes highly enhance the electromagnetic field. The FDTD simulations comparing the field enhancement of AuNRs and AuNSs are detailed in Supplementary Fig. S9.

The time axes in Fig. 1 b,c represent relative, rather than absolute, time. This normalization accounts for variations in the onset of neurotransmitter binding events, which differ due to diffusion dynamics. For instance, GABA might start showing detection events at 100 s; glutamate may at 80 s and dopamine at 150. To enable comparison across all events within a unified time frame, a 20 s time window was constructed by aligning traces to the start of the first detected event and plotting subsequent data points relative to that origin. For clarity, the resonance traces were vertically offset along the y-axis; these offsets do not affect step height measurements. The step amplitudes, which are key to the analysis, retain their true values and correspond directly to the y-axis scale (1 unit = 10 fm).

## Molecular interaction mechanisms elucidated through DFT simulations

To elucidate the molecular basis for the observed differences in detection signals, we performed density functional theory (DFT) calculations to probe the interaction of dopamine, glutamate, and GABA with gold surfaces under physiologically relevant conditions. The simulations employed the M06-2X functional, which offers accurate descriptions of noncovalent interactions and adsorption processes on metal surfaces[33]. A slab model of the gold (111) surface was constructed by cleaving and optimizing an $Au_{(22)}$ (111) facet to its lowest energy geometry[34]. Computational details for DFT have been described in supplementary section 8.

Interaction energies were computed for each neurotransmitter with the Au (111) surface both in the presence and absence of phosphate ions, which are prevalent under our experimental buffer conditions (pH < pK$_a$ of the amine groups). In the absence of phosphate, only molecules containing a free lone pair on the protonated amine (−NH$_2$) were found to weakly adsorb to the gold surface, yielding an interaction energy of ∼-23.35 kcal/mol, with chemical accuracy of ∼1 kcal/mol(Fig. 2b–c). However, in phosphate-buffered systems, phosphate anions ($HPO_4^{2-}/H_2PO_4^-$) chemisorb strongly onto Au (111), exhibiting an interaction energy of -114.03 kcal/mol, with chemical accuracy of ∼1 kcal/mol.

The adsorbed phosphate species serve as a hydrogen-bonding bridge between the gold surface and neurotransmitters bearing protonated amines (NH$_3^+$), significantly enhancing molecular adsorption (Fig. 2d–e). The calculated interaction energies in the prevalent of phosphate ions were found to be: dopamine (-282.58 kcal/mol), GABA (-144.13 kcal/mol), and glutamate (-112.35 kcal/mol). This trend mirrors our experimental measurements, where we observed permanent interactions (step signals) with all the mentioned neurotransmitters interacting with the optoplasmonic sensor in phosphate buffer. In addition to optoplasmonic single-molecule measurements, the interaction of GABA and glutamate with phosphate-anion-modified gold surfaces was experimentally verified using surface-enhanced Raman spectroscopy. (see supplementary Fig. S4).

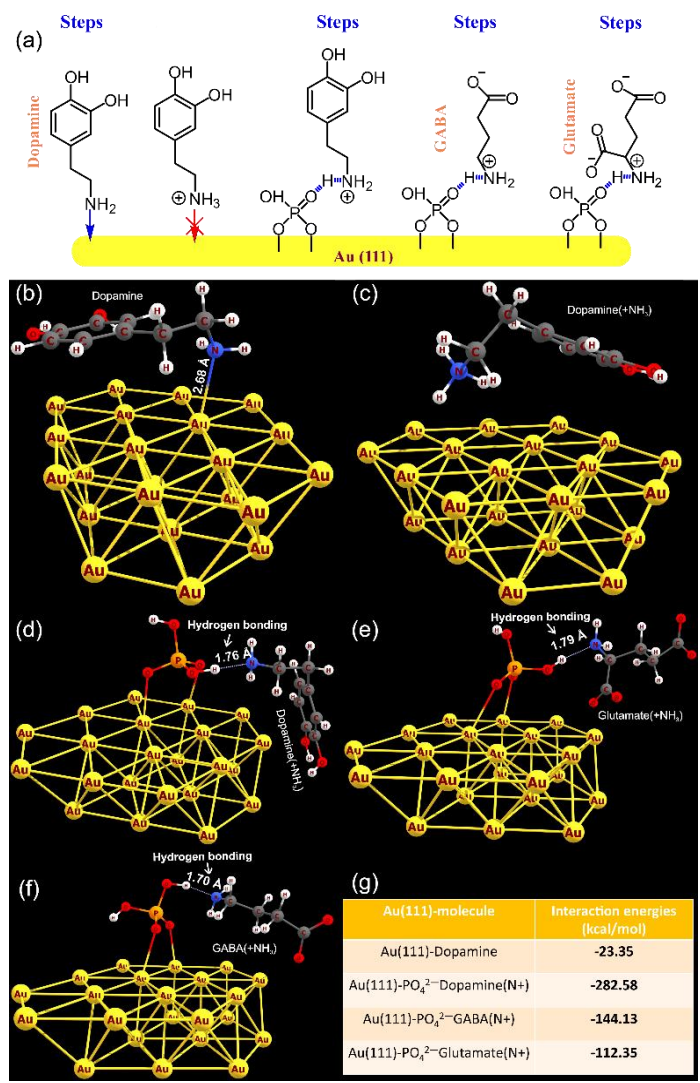

**Fig. 2 | DFT simulations of neurotransmitter interactions with the Au (111) surface .(a)** Schematics of type of interactions in Dopamine, GABA, and Glutamate with the Au (111) surface. **(b)-(f)** Optimized geometries of the neurotransmitters on Au (111), both in the absence and presence of phosphate molecules, obtained from density functional theory (DFT) calculations. **(g)** Table summarizing the calculated interaction energies of the neurotransmitters with the Au (111) surface.

## Modulating Sensor Behaviour via pH-Dependent Transient and Permanent Neurotransmitter Interactions

At physiological pH (7.4), step-like signals were exclusively observed, originating from the permanent interactions between the protonated amine groups of neurotransmitters and chemisorbed phosphate anions on the AuNP surface. These strong, irreversible interactions lead to molecular saturation of available binding sites, which in turn impedes subsequent analyte access and causes a progressive loss of sensor sensitivity. Despite this limitation, the platform retains its analytical utility through the detection of transient interactions, enabled by deliberate tuning of the experimental environment.

To explore this behaviour, we modulated the sensor environment by increasing the pH above the pKa of the neurotransmitter amines, thus shifting their protonation states and interaction profiles. In this higher pH regime (pH 10.6, 50 mM carbonate buffer), neurotransmitters displayed both step and spike signals, reflective of both, strong bonding and weak repulsive interactions, respectively. These transient spike signals, in particular, allow real-time monitoring of interaction kinetics between neurotransmitters and the sensor surface without signal saturation, which helps the sensor to capture signals for prolonged time, enabling to do the statistical analysis. (Supplementary Figs. S6–S8).

## GABA Interactions at High pH

We first investigated GABA, which, at pH 10.6, contains a deprotonated carboxylate and a neutral amine with a lone pair capable of forming coordinate bonds with gold surfaces. As shown in Fig. 3a (left), step-like resonant shifts result from coordinate covalent bond between GABA's amine group and AuNS tips. Concurrently, spike-like events occur due to electrostatic repulsion between GABA's carboxylate group (-COO⁻) and the negatively charged AuNS surface which are pre-chemisorbed with carbonate anions (-CO₃²⁻), as depicted in Fig. 3a (right).

Notably, simultaneous detection of step and spike signals occurred at specific concentrations (100 fM, 1 pM, and 10 nM). At other concentrations (e.g, 1 fM), spike activity diminished, likely due to GABA possessing only a single carboxylate group, limiting the extent of repulsive interactions with the carbonate anions on the AuNS surface. Fig. 3b presents the average step and spike heights for GABA at 1 pM and 10 nM. These averages were calculated by integrating the amplitude of all events within a concentration window and normalizing by total event count.

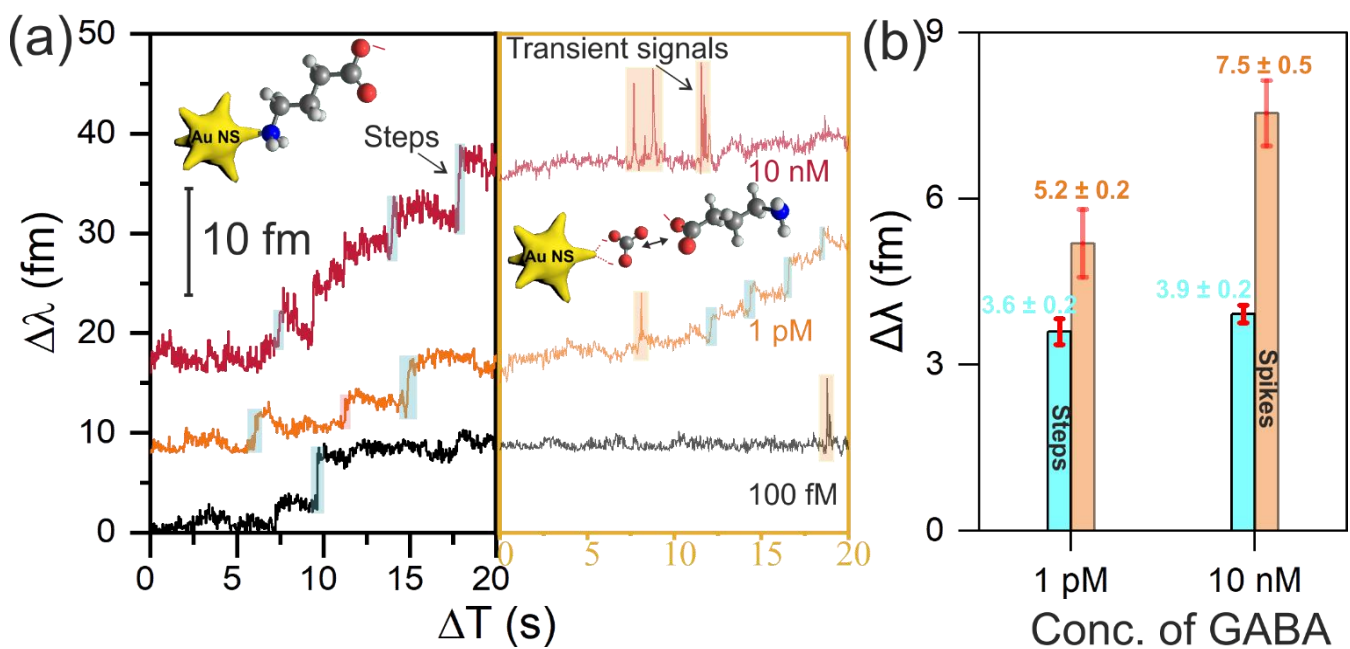

**Fig. 3. Interaction of GABA with AuNSs at pH 10.6. a,** Resonance shift time traces of GABA at concentrations of 100 fM, 1 pM, and 10 nM, recorded using gold nanostars (AuNSs) at pH 10.6. Data are shown over two distinct 20 s time windows. Discrete step and spike signals are observed. **b,** Average step heights and spike (peak) heights extracted from GABA signals at 1 pM and 10 nM concentrations.

## Glutamate Interaction Kinetics and Signal Profile

Glutamate demonstrated a more complex interaction profile, consistent with its dual carboxylate functionality. Like GABA, glutamate formed coordinate covalent bonds via its neutral amine at high pH, producing step events (Fig. 4a, left). However, the presence of two carboxylate groups significantly increased the frequency and intensity of spike signals, attributed to enhanced repulsive interactions with the carbonate anions on the negatively charged AuNS surface.

Fig. 4b illustrates representative resonance traces of glutamate at 10 aM, 100 fM, and 1 μM. Even at the attomolar level (10 aM), transient spike signals were reliably recorded, demonstrating the platform's ultra-sensitivity. Fig. 4c shows the average step and spike heights across these concentrations. At higher concentrations (e.g., 1 µM), molecules come into closer proximity and are more likely to form non-zwitterionic glutamate dimers through intramolecular hydrogen bonding. However, these dimers are relatively less stable due to the reduced number of neutral hydrogen bonds[35, 36]. Consequently, these dimers interact only transiently with the glutamate-bound sensor assembly, leading to increased spike heights at higher concentrations. GABA molecules may also form similar weak dimers that transiently interact with the GABA-bound microsphere–AuNS construct, resulting in a slight increase in spike height at 10 nM (Fig. 3b) compared to the average spike amplitude observed at 1 pM. While computational studies have explored these molecular interactions, no experimental evidence has yet confirmed dimerization of neurotransmitters. This highlights the significance of optoplasmonic WGM sensors, which are capable of detecting such transient molecular events with high sensitivity (see supplementary Fig. S5).

These observations are in excellent agreement with our theoretical modelling: glutamate, which possesses a greater molecular polarizability than GABA, consistently induced larger average resonance shifts across all tested concentrations. This aligns with predictions that the resonance shift is directly proportiona[37] to the real component of the polarizability tensor of the interacting analyte.

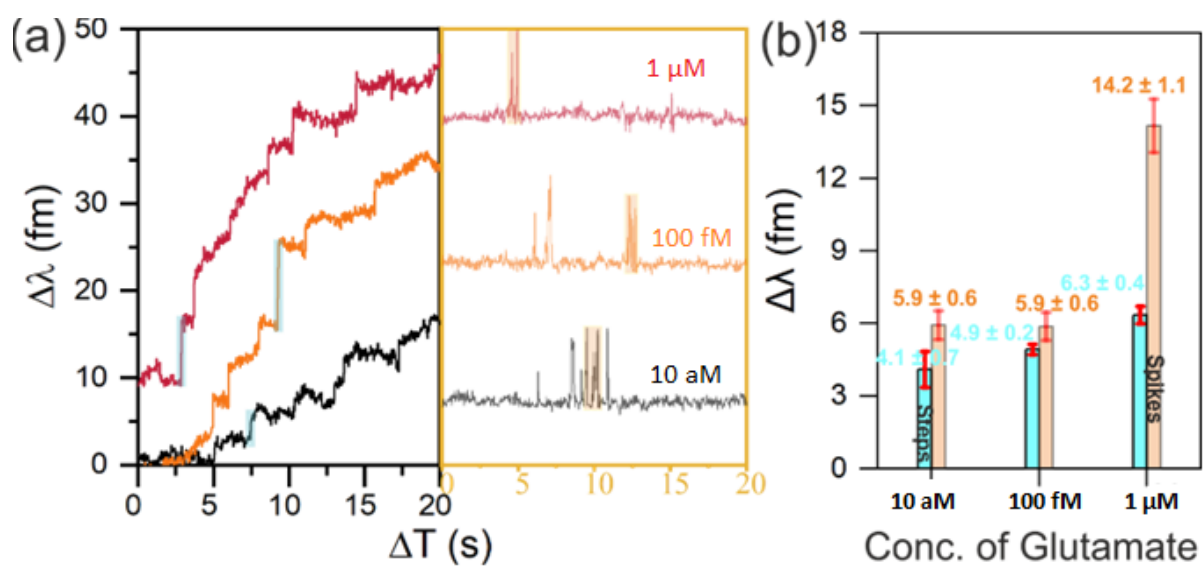

**Fig. 4 | Interaction of glutamate with AuNSs at pH 10.6** . **a**, Resonance shift time traces recorded at glutamate concentrations of 10 aM (blue), 100 fM (orange), and 1 µM (red) at pH 10.6, shown over two distinct 20 s time windows. Step and spike events are highlighted i. **c**, Average step height and average spike (peak) height extracted from resonance data at 10 aM, 100 fM, and 1 µM glutamate concentrations

## Selective differentiation of neurotransmitters with functionalized optoplasmonic WGM sensors

Thus far, we have demonstrated that optoplasmonic WGM sensors offer exceptional sensitivity in detecting neurotransmitters under varying environmental conditions. The recorded sensor signals exhibit strong correlation with molecular polarizability but are also modulated by the spatial distribution of local electromagnetic fields at the binding interface. This interplay introduces statistical dispersion across detection events and complicates analyte discrimination based solely on signal magnitude, especially with AuNSs, although it improves signal amplitude. To address this challenge, we employ the widely studied and broadly applicable class of nanoparticles-AuNRs, deposited on the microresonator surface, and functionalized with a receptor molecule, enabling the extraction of neurotransmitter-specific kinetic fingerprints through molecular discrimination.

Here, we introduce a molecular interface engineering approach using 3-mercaptopropionic acid (3-MPA), a bifunctional linker that modulates surface chemistry without contributing directly to the resonant signal. By the judicious choice of the environmental conditions based on the $pK_a$ values of neurotransmitter functional groups, we attain a label-free and selective differentiation of GABA and glutamate, which are chemically similar in structure but entirely different functional characteristics.

Differentiating between GABA and glutamate is of significant neuroscientific relevance due to their opposing roles in synaptic transmission and their involvement in numerous psychiatric and neurological disorders [38] Within the cortical and synaptic environments, GABA and glutamate exist at concentrations ranging from 0.1–1 µM and 0.5–1 µM, respectively, levels at which traditional techniques such as microdialysis [38] and SERS [39] struggle to achieve high selectivity and sensitivity. To distinguish GABA from glutamate, AuNP-functionalized WGM microspheres were incubated with 10 mM 3-MPA at pH 7.7 for 30 minutes to enable thiol-mediated chemisorption. Post-incubation, unbound molecules were removed, and the sensing chamber was flushed with carbonate buffer at pH 9.7, carefully selected based on the protonation states of the functional groups in each neurotransmitter. At this critical pH 9.7, GABA exists with a protonated amine group (pH 9.7 < pKa 10.5) and a carboxylate anion (pH 9.7 > pKa 4.03), while glutamate possesses two carboxylate anions (pH 9.7 > pKa 2.19) and an amine group with a lone pair (pH 9.7 ≈ pKa 9.67).

Fig 5a schematically depicts the interaction modalities. The protonated amine of GABA forms stable hydrogen bonds with the carboxylate group of 3-MPA, facilitated by GABA's compact structure and favourable orientation. These interactions generate discrete, step-like resonant shifts, recorded at both 10 nM and 1 µM concentrations (Fig. 5b).

In contrast, glutamate, with two carboxylate groups and no significant hydrogen bonding capability under these conditions, interacts repulsively with the negatively charged 3-MPA-functionalized surface. These interactions manifest as transient spike-like signals, also consistently observed at 10 nM and 1 µM concentrations (Fig. 5c).

Equimolar mixtures of GABA and glutamate (10 nM each) were introduced to the PDMS chamber to demonstrate distinguishability capability of the hybrid sensor. The following signals (Fig. 5d) observed both permanent step events (GABA) and transient spikes (glutamate), clearly validating the simultaneous and independent single-molecule detection of both neurotransmitters in mixed environment conditions.

This study introduces a novel molecular differentiation strategy using pH- and interface-tuned optoplasmonic WGM sensors capable of distinguishing structurally similar neurotransmitters. By leveraging specific interaction geometries, hydrogen bonding for GABA and electrostatic repulsion for glutamate, mediated by 3-MPA, we establish a robust, label-free sensing approach with single-event resolution and high sensitivity (demonstrated down to 10 nM).

The ability to discriminate GABA and glutamate in real-time at physiologically relevant concentrations offers profound implications for probing neurotransmitter dynamics in brain research, psychiatric diagnostics, and neurodegenerative disease monitoring. Furthermore, the modularity of this approach enables facile extension to other biomolecules with subtle structural differences.

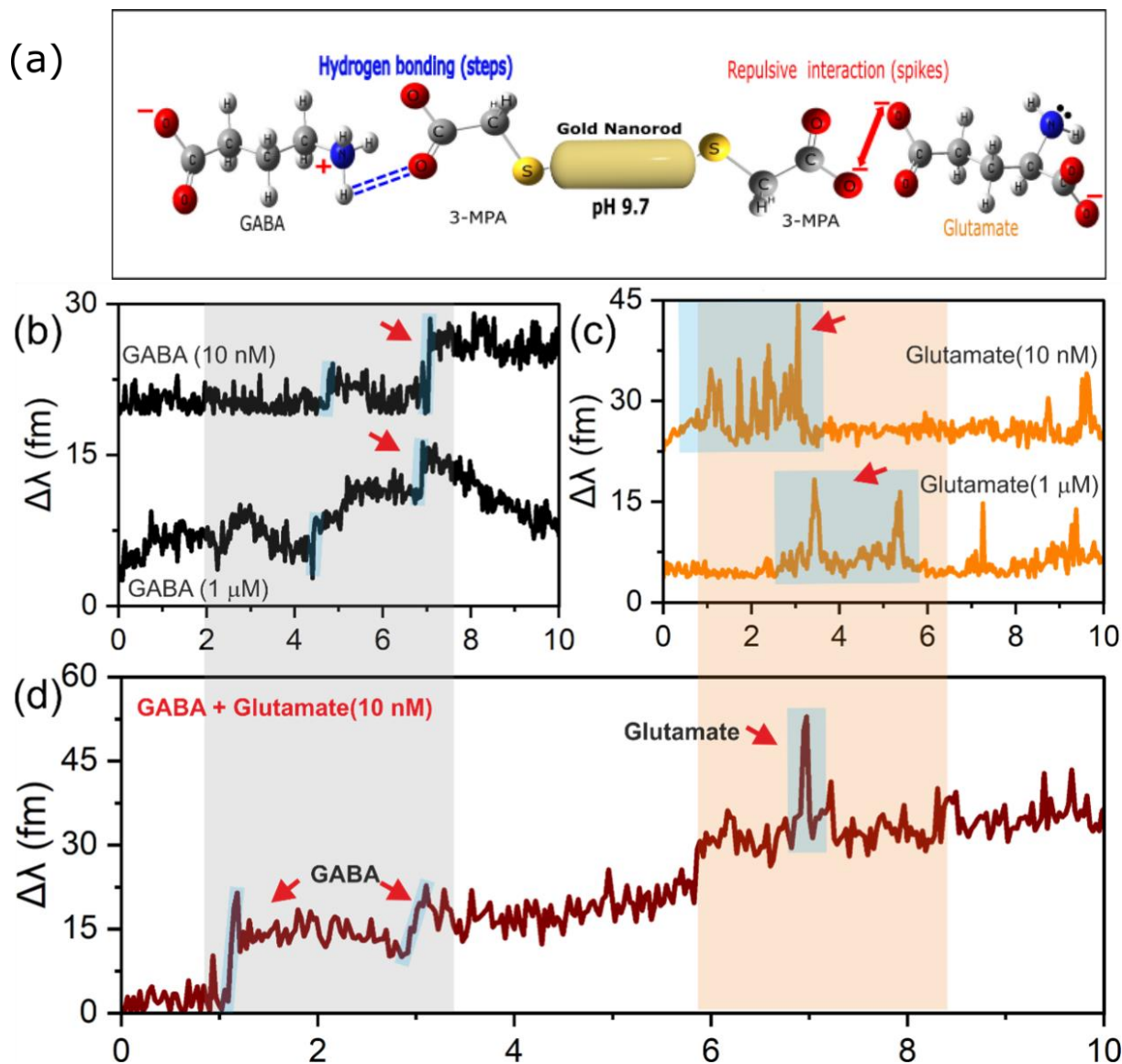

**Fig . 5. Distinguishing GABA and glutamate using 3-MPA-functionalized AuNRs a,** Schematic illustration of the interactions between GABA and glutamate with Au nanorods (AuNRs) functionalized with 3-mercaptopropionic acid (3-MPA). GABA forms hydrogen bonds with 3-MPA, while glutamate experiences repulsive interactions. **b,** Resonance shift time traces for GABA at concentrations of 10 nM and 1 μM, showing distinct stepwise responses. **c,** Transient spike responses observed for glutamate at 10 nM and 1 μM, attributed to repulsive interactions with 3-MPA functionalized surface. **d,** Mixed solution of GABA and glutamate (each at 10 nM) reveals both permanent steps (GABA) and transient spikes (glutamate), demonstrating the capability of the optoplasmonic WGM sensor to distinguish between the two neurotransmitters. All the data are shown over a relative 10 s time window.

## Discussion and Conclusion

The optoplasmonic WGM sensing platform presented in this work constitutes a substantial advancement in label-free single-molecule detection technology, with promising applications in neurochemical sensing. By leveraging the exquisite sensitivity of high-Q WGM microresonators enhanced with plasmonic nanoparticles, we have achieved single-molecule detection of neurotransmitters at concentrations spanning from 10 aM to 1 μM, a range that encompasses physiologically relevant concentrations in the synaptic cleft after diffusion, receptor binding, and transporter-mediated reuptake. Importantly, our platform offers millisecond temporal resolution, matching the timescale of synaptic processes and enabling real-time monitoring of rapid neurochemical dynamics.

Our comparative analysis of AuNRs and AuNSs as plasmonic enhancement elements revealed that the three-dimensional star morphology significantly outperforms nanorods, with detection enhancements of 40.6%, 34.6%, and 51.8% for GABA, glutamate, and dopamine, respectively. This improvement stems from the multiple sharp tips of AuNSs that generate stronger and more numerous electromagnetic hot spots. The observed increase in on-rate for neurotransmitter interactions with nanostars further confirms their superior sensing efficiency. These findings provide valuable design principles for future optoplasmonic sensors targeting small biomolecules and suggest that nanostructures with even more pronounced field confinement could further enhance detection capabilities.

The observed correlation between resonant shift magnitudes and molecular polarizabilities demonstrates that our platform probes intrinsic electronic properties of analyte molecules. The ability to extract molecular information beyond mere presence/absence detection represents an important advancement toward multidimensional optical biosensing.

Extensive density functional theory (DFT) calculations reveal that phosphate anions, present as buffer components, mediate the interaction between neurotransmitters and the AuNP surface via hydrogen-bonding bridges. The calculated interaction energies between phosphate anions and the protonated amine groups of the neurotransmitters are consistent with experimental observations. These interactions were observed as distinct step-like signals in the measurements, corresponding to individual binding events.

Most importantly, our discovery of pH-dependent interaction dynamics, and their role in the selective differentiation of neurotransmitters, such as GABA and glutamate, addresses a fundamental challenge in neurochemical sensing. By functionalizing the sensor surface with 3-MPA and operating at pH 9.7, we demonstrated unambiguous discrimination between GABA and glutamate based on their characteristic binding signatures, which were permanent step signals for GABA versus transient spike signals for glutamate. This differentiation capability was maintained even in equimolar mixtures, highlighting the potential for multiplexed detection in complex biological environments. This accomplishment makes WGM sensing stand out from the conventional high-fidelity techniques like HPLC/MS, which are capable of discriminating neurotransmitters, but lack single-molecule sensitivity and have comparably poor detection limits.

The WGM sensograms trace the kinetics of neurotransmitters as they interact with the biosensing fields. The transient signals (spikes) are particularly important for probing rate kinetics and enable continuous monitoring without sensor saturation unlike the step signals, which are associated with binding events. Despite these advances, several challenges remain. For dopamine, polymerization at pH > 7 prevents single-molecule spike signal detection and creates a thin coating that interferes with optical coupling. Additionally, at very low concentrations (10 aM, 1 fM), the limited number of spike events for GABA and glutamate resulted in statistical analyses that were not ideally smooth. Future work could address these limitations through alternative protocols, more favourable environmental conditions (pH > 10.6), or nanoparticles with enhanced electric field localization properties.

We also note that while 3-MPA functionalization enables selective differentiation of neurotransmitters, the chemisorption process introduces slightly higher noise in the resonance traces, necessitating operation at higher concentrations (10 nM and 1 μM) for reliable signal discrimination. This represents a trade-off between selectivity and ultimate sensitivity that future refinements might overcome through alternative surface functionalization strategies or improved signal processing algorithms.

From a photonics perspective, this work demonstrates how the integration of distinct optical phenomena, whispering gallery modes and localized surface plasmon resonances, can create hybrid sensing modalities with performance characteristics unattainable by either approach alone. The coupling between the high-Q resonator and plasmonic nanostructures enables both extreme sensitivity and molecular specificity through electromagnetic field enhancement and controlled surface chemistry. This combination surpasses existing neurotransmitter detection techniques such as FRET-based methods, electrochemical approaches, and conventional optical fiber sensors, which are typically ensemble-based with detection limits around 10 nM.

Looking ahead, our optoplasmonic WGM sensing technique could be used in conjunction with methods, such as microdialysis or cerebrospinal fluid sampling, for clinical neuroscience research. However, the presence of non-target molecules in the extracellular fluid could introduce additional noise into the WGM sensograms, compromising the specificity and sensitivity of the analysis. To mitigate this issue, an intermediate filtering step could be used to isolate the neurotransmitter molecules from the surrounding fluid. In the more immediate future, our sensor could be used to monitor the release of neurotransmitters in a controlled environment, such as cultured hippocampal neurons. The absence of extracellular fluid in a cell culture medium simplifies the analysis and allows for precise calibration of stimulation parameters to trigger release without cellular damage. This approach would facilitate real-time monitoring of neurotransmitter release, providing detailed insights into the dynamics of neuronal communication. Additional research directions include studying neurotransmitter interactions with lipid bilayers coated on the microresonator, which could shed light on alternative pathways for neurotransmission based on the lipophilic or lipophobic nature of these signalling neurotransmitters.

The work bridges advanced photonic technologies with critical challenges in neuroscience, potentially enabling new investigations into neurochemical signalling with unprecedented precision. The platform we have developed and calibrated provides a powerful tool for characterizing neurotransmitters at the single-molecule level, laying the foundation for advanced opto-synaptic interfaces capable of real-time brain activity analysis, and with further on-chip integration, of *in vivo* neural probing applications [40]. Beyond neurotransmitters, the sensing principles demonstrated here could be extended to other small biomolecules of biomedical importance, establishing optoplasmonic WGM sensing as a versatile platform for single-molecule analytics.

## Methods

### 1 Optoplasmonic WGM Experimental Setup:

WGMs were excited in microspherical resonators using a continuous-wave laser with a nominal wavelength of $\lambda_o$ = 780 nm. The output beam was collimated via a fiber collimator and its diameter adjusted using a variable beam expander. A half-wave plate and a quarter-wave plate were placed in the input and output optical paths, respectively, to selectively excite transverse electric (TE) and transverse magnetic (TM) modes. Transverse electric mode has been used in the experiment to excite the longitudinal plasmon resonances of the CTAB-AuNPs.

The laser beam was focused onto the surface of a high-refractive-index glass prism (N-SF11, n ≈ 1.77) using converging lenses to facilitate evanescent coupling via frustrated total internal reflection (FTIR). Microspherical resonators (diameter: 85–90 μm), fabricated by melting the tapered end of SMF), were positioned at the prism interface to enable efficient mode excitation. The spot size and alignment were optimized to maximize coupling into fundamental WGM modes[41].

Transmitted light was collected by a photodiode placed after a pair of converging lenses. A variable iris was used to spatially filter the beam and suppress unwanted scattering. The coupling region was imaged using a microscope assembly comprising a 10× objective lens and a CMOS camera to facilitate real-time alignment and monitoring.

A polydimethylsiloxane (PDMS) chamber (~300 μL volume, V-shaped geometry) was affixed to the prism surface to contain liquid samples and buffer solutions. The chamber was manually cleaned and refilled using a pipette between experiments. A thick glass coverslip was bonded to the rear face of the PDMS (stick to the prism) to enclose the chamber, and a thermoelectric cooler (TEC) was mounted behind the coverslip to support the chamber and to provide active temperature regulation when required. All the experiments were performed at room temperature 24°C. Fine control of the resonator–prism coupling gap and beam alignment was achieved using an XYZ piezoelectric translation stage, enabling optimization of Q-factor and coupling efficiency.

### 2 Functionalization of AuNPs on the microcavity surface:

The cetyltrimethylammonium bromide (CTAB) capped gold nanoparticles (AuNPs) were electrostatically immobilized onto the silica microsphere surface under highly acidic conditions (24 mM HCl, pH 1.6) within the PDMS sample chamber. To prevent aggregation, AuNPs were sonicated for 10 min prior to use and diluted in a 1:4 ratio to limit the number of particles binding to the resonator. Electrostatic attraction between the negatively charged silica microsphere surface and the positively charged CTAB layer on the AuNR facilitated attachment of nanoparticles to the microsphere.

Real-time monitoring of AuNP binding events was performed by tracking rapid shifts in both the resonance wavelength and the full width at half maximum (FWHM) of the WGM signal. At a nanoparticle concentration of 2.5 pM, typically 5–6 AuNPs attached per microsphere. Following attachment, the chamber was rinsed twice with ultrapure water to prevent additional nanoparticle binding, which can degrade the quality factor (Q) of the resonator. Also, after the rinsing the chamber was subsequently incubated with same HCL buffer (24 mM, PH 1.6) for at least 10 mins, to ensure the stable immobilization of the nanoparticles to the microsphere.

Given the sensitivity of single-molecule detection, all solutions were freshly prepared immediately before each experiment. Buffers were filtered using 0.2 µm Minisart NML Plus glass-fiber, surfactant-free cellulose acetate syringe filters (Sartorius) and vortexed before use. Neurotransmitter stock solutions (100 mM) of γ-aminobutyric acid (GABA), glutamate, and dopamine were prepared using the same buffer system (phosphate or carbonate) as used in the experiments. Working concentrations were serially diluted from these stocks.

Measurements began at 10 aM neurotransmitter concentration, followed by sequential increases to 1 fM, 100 fM, 10 pM, 10 nM, and 1 µM. All the concentrations were recorded for the same time-20 mins. Also, each neurotransmitter was tested with independent sensor assembly (microsphere-AuNP) ensuring the specificity of the neurotransmitter and reproducibility of the signals.

## 3 Data collection:

The WGM transmission spectra were recorded and analyzed in real time using a custom-built LabVIEW program 'BaaskeDAQ' which implements a bespoke centroid-based algorithm. This method enabled accurate estimation and tracking of the resonance wavelength and FWHM across time-traces.

The resulting time traces were further processed using a custom graphical user interface, *Data Analysis*, developed in MATLAB. This tool was specifically designed to identify, extract WGM signal features, enabling precise characterization and analysis of single-molecule events.

### 3a Spike height extraction:

External perturbations, including pressure fluctuations, thermal variations from the sample chamber, and mechanical shifts induced by the thermoelectric cooler (TEC), can cause slow drifts in the WGM resonance position during measurements. To correct for these baseline fluctuations, a detrending procedure was implemented.

A first-order Savitzky–Golay filter [42] with a window length of 101 data points was applied as a low-pass filter to extract the slow-varying background. The filtered trace was subsequently subtracted from the original resonance data to yield the detrended signal, isolating rapid shifts attributable to single-molecule interactions.

Transient resonance shifts, or "spikes," are attributed to temporary interactions between neurotransmitter molecules and the AuNP-modified microsphere surface. These spike events resemble noise, but they are recognizable by their greater amplitude. To ensure robust signal discrimination, only events exceeding a threshold of 3σ were classified as WGM spike signals, where σ denotes the standard deviation of the background noise.

To determine σ, the time trace was segmented into windows of $N$ points, where $N$ ranged from 10 to 1,000 depending on the data acquisition (DAQ) frequency. The background noise level was estimated as the minimum standard deviation computed across all windows. The typical value of σ was found to range between 0.5 and 0.7 fm, can vary according to the ionic conditions and chemical composition of the buffer.

### 3b Step height extraction:

Permanent binding interactions, manifested as discrete step-like shifts in the WGM resonance, were identified through a semi-manual procedure. Step events were first located by visual inspection of the time trace, followed by quantitative analysis.

The magnitude of the step signals was calculated from the WGM resonance traces by performing two linear fits on 100 data points, one just before and one just after the occurrence of a step event. The amplitude of the step signals was determined from the vertical offset between these two fitted lines at the location of the step event, which was manually identified. Similar to spike height extraction, only step signals that exceed $3\sigma$ are considered WGM step signals, where $\sigma$ represents the standard deviation of the background.

## Data availability

All data supporting the findings of this study are available within the article and its supplementary information.

## Acknowledgements

The authors acknowledge the funding from UKRI Engineering and Physical Sciences Research Council [EP/R031428/1 and also EP/T002875/1]. A.K.A gratefully acknowledges the support provided by the Research Training Support Grant (RTSG) from the University of Exeter. A.K.A thank C. Jones, K. Perera and A. Attenborough for the insightful discussions and valuable feedback on the paper.

## Author contributions

F.V and J.X conceived the idea and planned the project. J.X built the single neurotransmitter detection experimental set-up, trained and supervised A.K.A. A.K.A performed the experiments and analysed the data. S.P carried out DFT simulations and SERS measurements. E.Z and J.X performed the FDTD and FEM simulations. E.Z performed GPAW simulations with the supervision of MW. F.V supervised the overall project. A.K.A wrote the manuscript, and all authors contributed to finalizing the content.

# Supplementary information

## Probing the single neurotransmitters with the WGM microcavity-hybridized plasmonic nanospiked antennas

A. K. Arunkumar [1, *] , E. Zossimova[1, 2] , M. Walter [2] , S. Pedireddy[1] , J. Xavier[1, 3, *] , F. Vollmer,[1]

## 1 UV-Vis absorption spectra of plasmonic nanoparticles

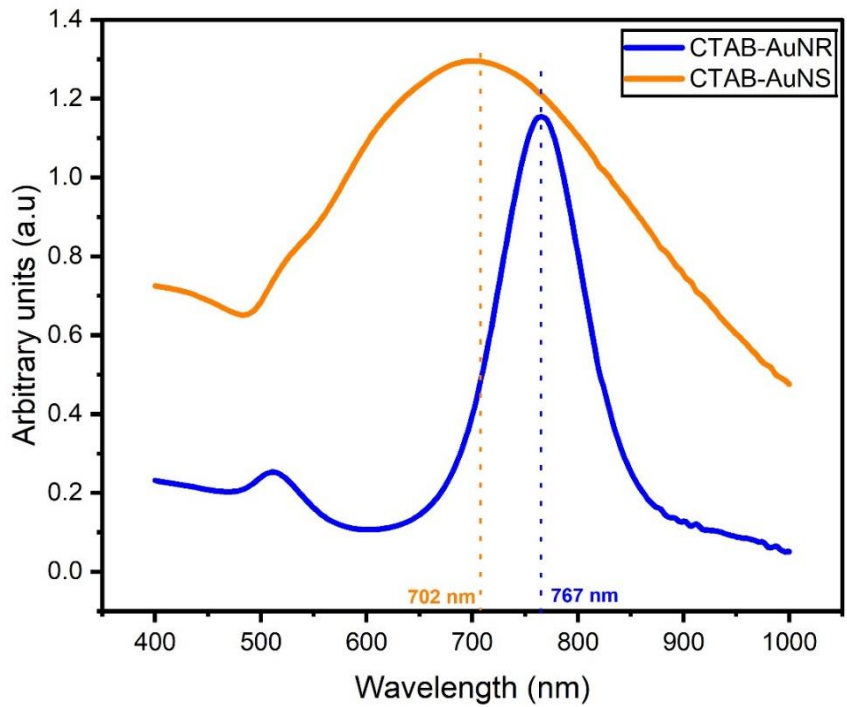

**Fig. S1 : Plasmon resonance spectra of CTAB-stabilized gold nanorods (AuNRs, blue) and nanostars (AuNSs, orange), showing longitudinal peaks at 767 nm and 702 nm, respectively.** Both peaks are blue-shifted relative to the 780 nm excitation laser, facilitating efficient coupling to whispering-gallery modes with reduced optical loss.

**Fig.S1** presents the UV–vis absorption spectra of CTAB-stabilized gold nanorods (AuNRs) and nanostars (AuNSs), both exhibiting plasmon resonance peaks blue-shifted relative to the 780 nm excitation laser. These nanoparticles with spectrally blue detuned plasmon resonance were deliberately selected, as their immobilisation on the microresonators induces a red shift. This shift brings the plasmon resonance closer to the Whispering Gallery Mode (WGM) probing wavelength, thereby providing on-resonant coupling conditions.

[1] Department of Physics and Astronomy, Living Systems institute, University of Exeter, Exeter, UK. [2]Freiburg Center for Interactive Materials and Bioinspired Technologies (FIT), University of Freiburg, Freiburg, Germany. [3]Currently with SeNSE, Indian Institute of Technology, Delhi, India. E-mail : a.kakkanattu-arunkumar2@exeter.ac.uk; j.xavier@exeter.ac.uk; f.vollmer@exeter.ac.uk

## 2 Plasmonic Enhancement and Coupling in WGM Sensors

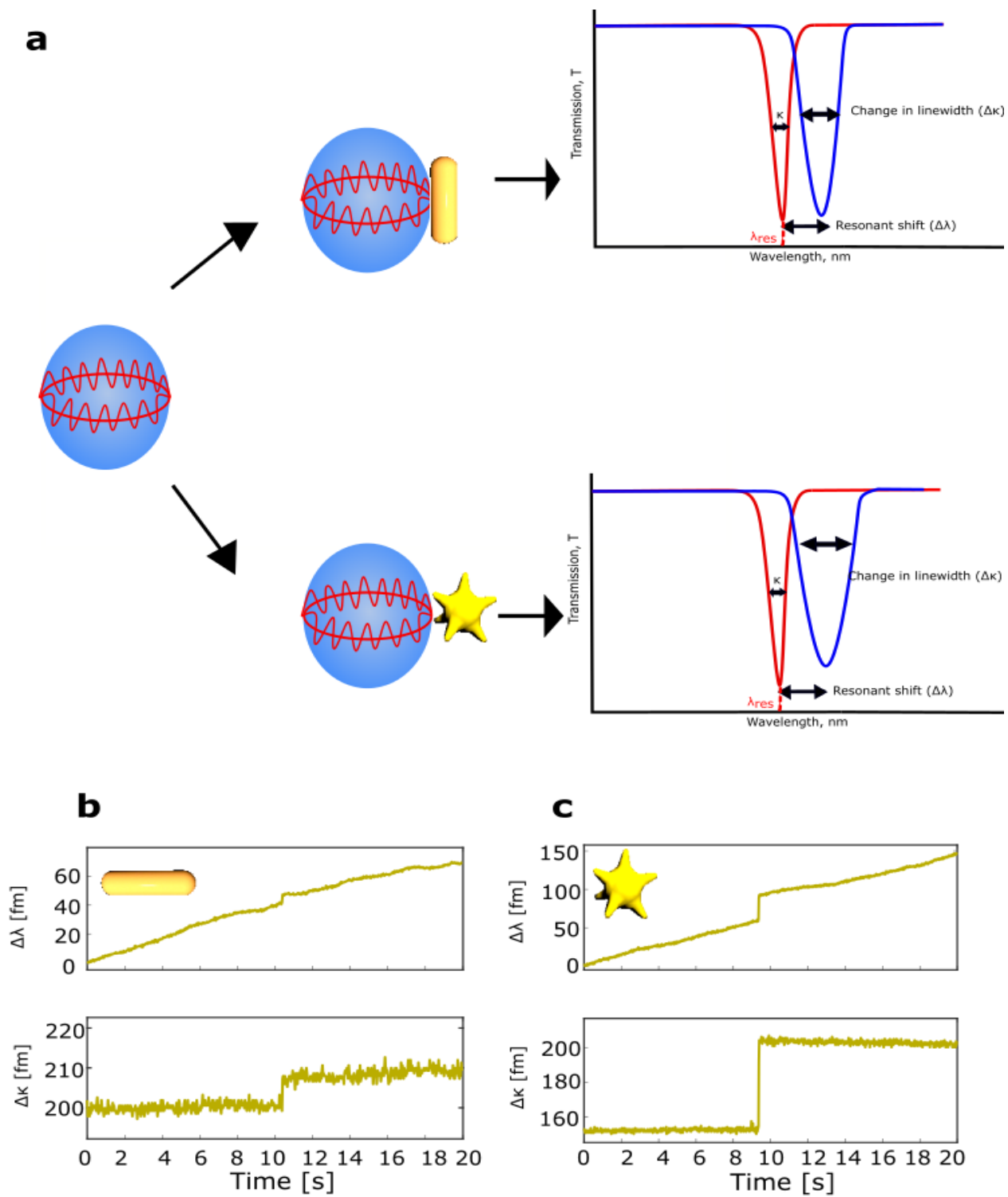

**Fig. S2 : Interaction of CTAB-stabilized gold nanoparticles with a WGM microresonator.** *a,* Schematic showing the binding of CTAB-coated gold nanorods (AuNRs) and nanostars (AuNSs) to the surface of a silica microsphere, alongside representative changes in the optical transmission spectra. *b,* Time-resolved resonance wavelength shift and linewidth broadening induced by the binding of a single AuNR. *c,* Corresponding spectral changes following the attachment of a single AuNS. Discrete step-like features in both cases indicate individual nanoparticle binding events.

To enhance the sensitivity of the WGM sensor, plasmonic gold nanoparticles (AuNPs) were selectively attached to the surface of the microsphere resonator. The coupling between the localized surface plasmon resonance (LSPR) of the AuNPs and the WGMs is governed by both the intensity of the evanescent field and the relative orientation of the nanorod with respect to the polarization of the WGM field. The evanescent field intensity is maximized near the equatorial plane of the microsphere, where the WGM is confined and circulates along the resonator's curved surface.

The overall sensitivity of the sensor is determined by a balance between the resonance quality factor (Q-factor) and the optical mode volume. Whilst plasmonic nanoparticles incorporated on the microsphere concentrate the fields to nanoscale volumes, they also introduce optical losses that degrade the Q-factor of the sensor. Therefore, there exists a trade-off between optimizing the Q-factor and mode volume when adding plasmonic nanoparticles. This trade-off is particularly relevant for applications requiring high spatial confinement, such as the detection of low-concentration neurotransmitters. As shown in Fig. S2 a,b, the typical linewidth broadening resulting from the attachment of a single AuNR is approximately 10 femtometres (fm), whereas the attachment of a single nanostar leads to a linewidth increase of ~40 fm.

## 3. The statistical analysis of permanent interactions of neurotransmitters

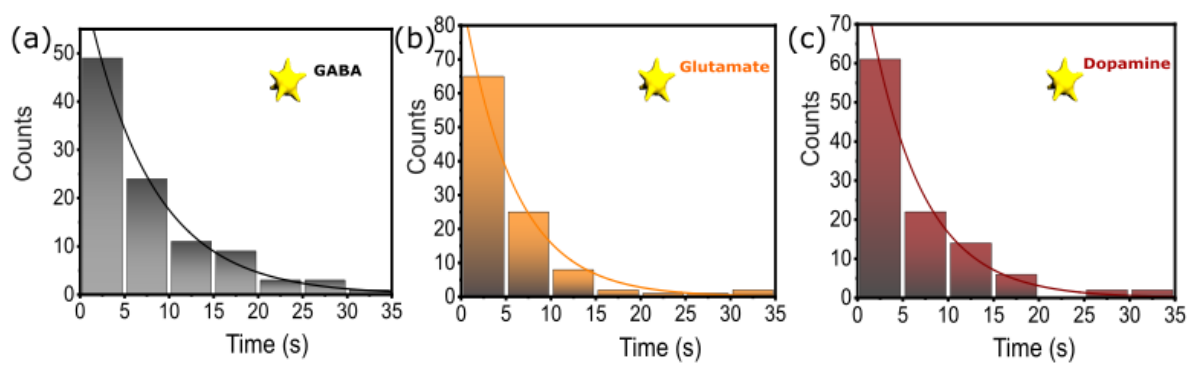

**Fig. S3 : Statistical analysis of single-molecule neurotransmitter binding events**. Histograms of time intervals between consecutive binding events for *a,* GABA, *b,* glutamate, and *c,* dopamine. The distributions follow single-exponential decay, consistent with Poissonian binding kinetics. Each dataset was obtained using a separately assembled whispering-gallery mode (WGM) microsphere sensor functionalized with CTAB-stabilized gold nanostars (AuNSs) tailored to the respective neurotransmitter.

## 4 SERS signals from amine-gold interaction of neurotransmitters

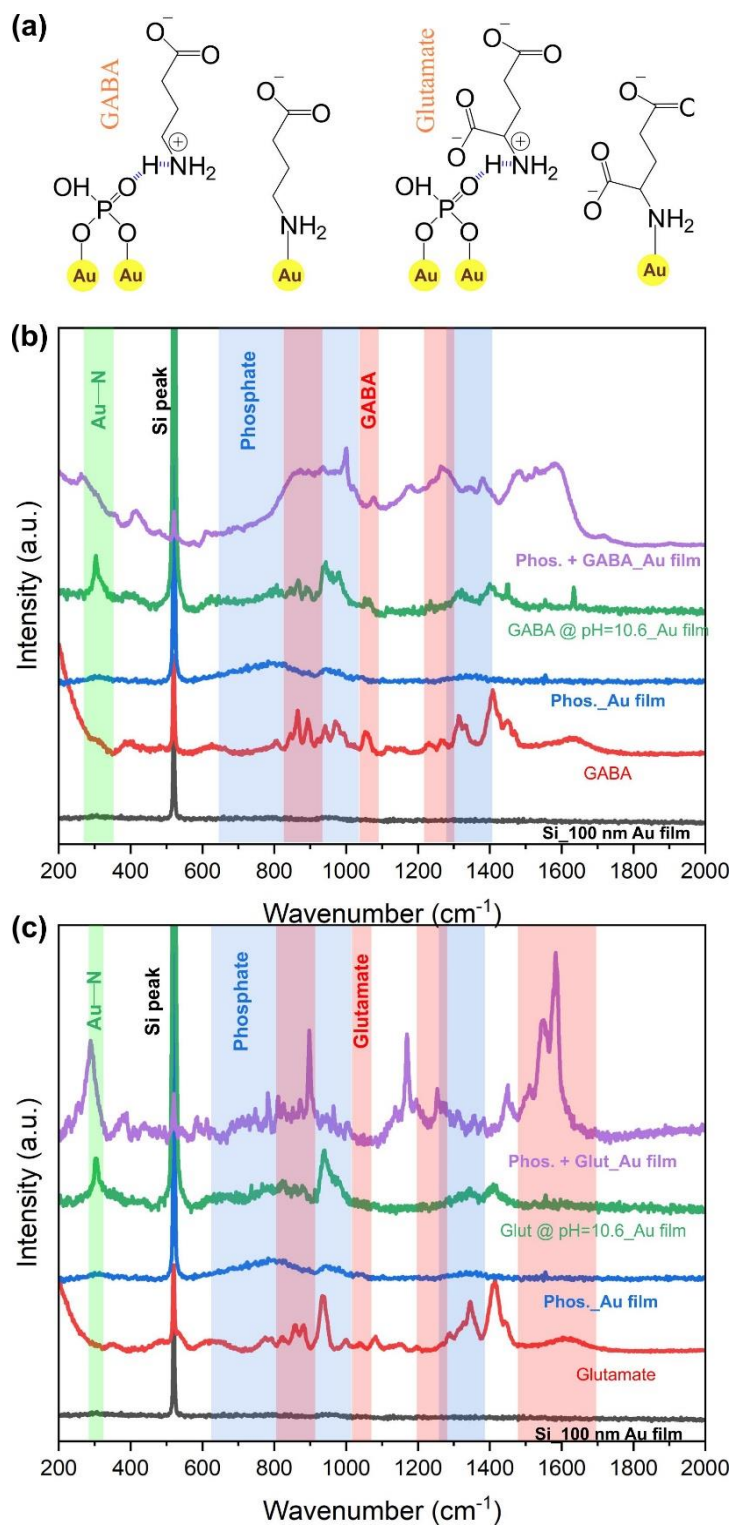

**Fig. S4: Schematics of design of SERS measurements.** (b)SERS spectra of Si_Au film, GABA droplet on Au film, phosphate chemisorbed on Au film, GABA molecules adsorbed on to Au film at pH=10.6, and GABA molecules interacting Au film containing adsorbed phosphate ions respectively. (c)SERS spectra of Si_Au film, glutamate droplet on Au film, phosphate chemisorbed on Au film, glutamate molecules adsorbed on to Au film at pH=10.6, and glutamate molecules interacting Au film containing adsorbed phosphate ions respectively.

A 100-nm-thick gold film was thermally evaporated onto a silicon substrate to serve as a SERS platform mimicking the Au(111) facets present on Au nanorod tips. The Au-coated Si substrates were cleaned using $O_2$ plasma to remove organic contaminants. Substrates were then incubated in 25 mM phosphate buffer (pH 7.4) for 30 min in Eppendorf tubes, followed by rinsing with deionized water. Subsequently, the substrates were incubated in 1 mM aqueous solutions of γ-aminobutyric acid (GABA) and glutamate for 30 min, rinsed with deionized water, and dried under a stream of $N_2$ gas. For bulk Raman spectra, 10 mM solutions of GABA and glutamate were drop-cast onto the Au-coated substrates and dried under vacuum prior to measurement. Raman spectra were acquired using a Renishaw in Via confocal microscope equipped with a 785 nm excitation laser and a long working distance 50× objective (NA = 0.45).

## 5 Glutamate monomers and dimers at 10 aM

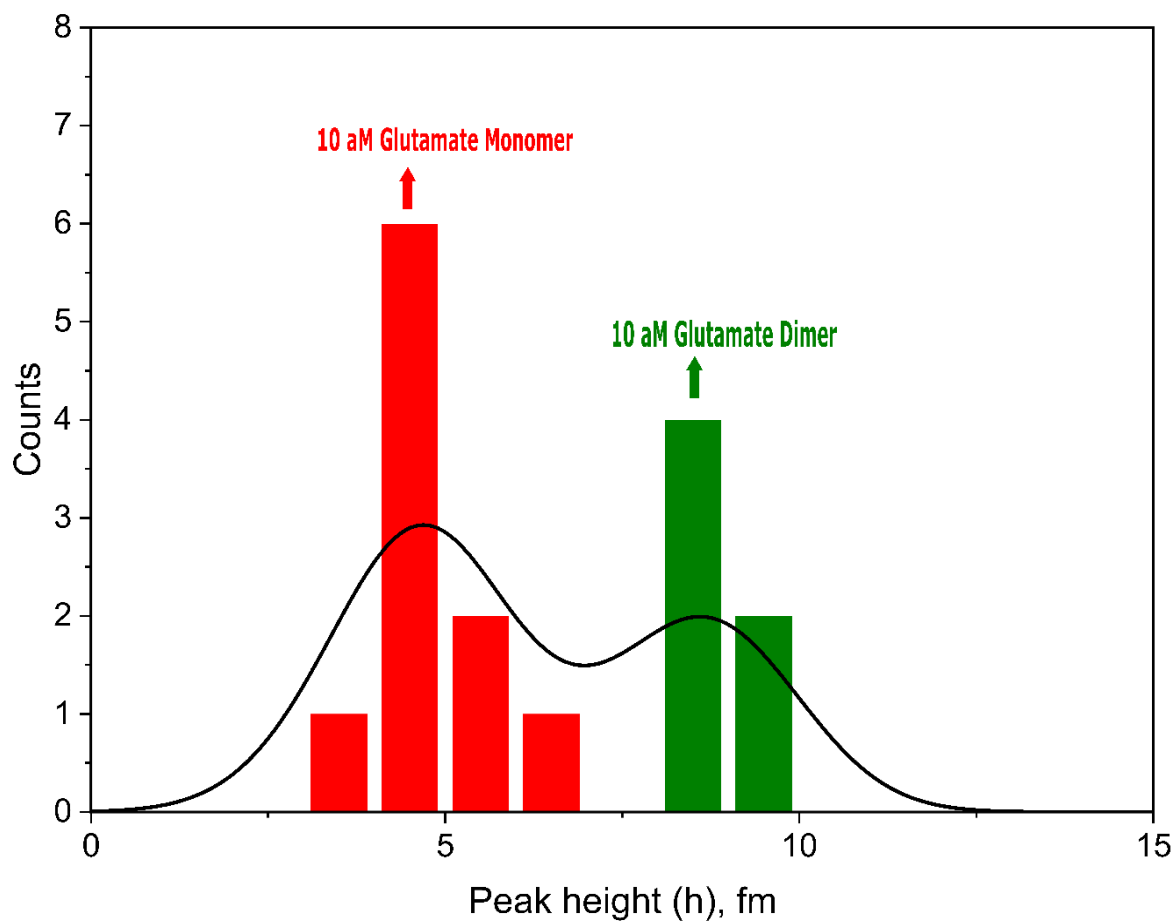

**Fig. S5 : Peak (spike) height distribution of 10 aM glutamate in pH 10.6**. Train of signals obtained from 10 aM glutamate monomer (Red) dimer (green).

The interaction of monomers and dimers with AuNSs was detected as two distinct signal trains in single-shot measurements at ultra-low concentrations (10 aM), characterized by lower-magnitude spikes for monomers (red) and higher-amplitude spikes for dimers (green). However, this distinction diminished at higher concentrations, likely due to molecular interactions from greater number of molecules and competition for binding at the limited available sites on the AuNSs, contrasting with the more specific interactions observed at lower concentration with more available interaction sites.

## 6 The interaction kinetics of neurotransmitters

Binding events were analyzed statistically by constructing histograms of the time intervals between consecutive spikes and the corresponding dwell times. When binding events occur stochastically and independently, and the rate of occurrence varies linearly with analyte concentration, the process can be treated as a single-molecule reaction. Under these conditions, the event follows Poissonian statistics, and mathematically represented as

$$P(n, R, \Delta t) \ = \ \frac{(R\Delta t)^n}{n!} e^{-R\Delta t} \qquad\qquad (S1)$$

Where R is the mean-event rate and $\Delta t$ is the time separation between two consecutive events (spikes) in seconds.

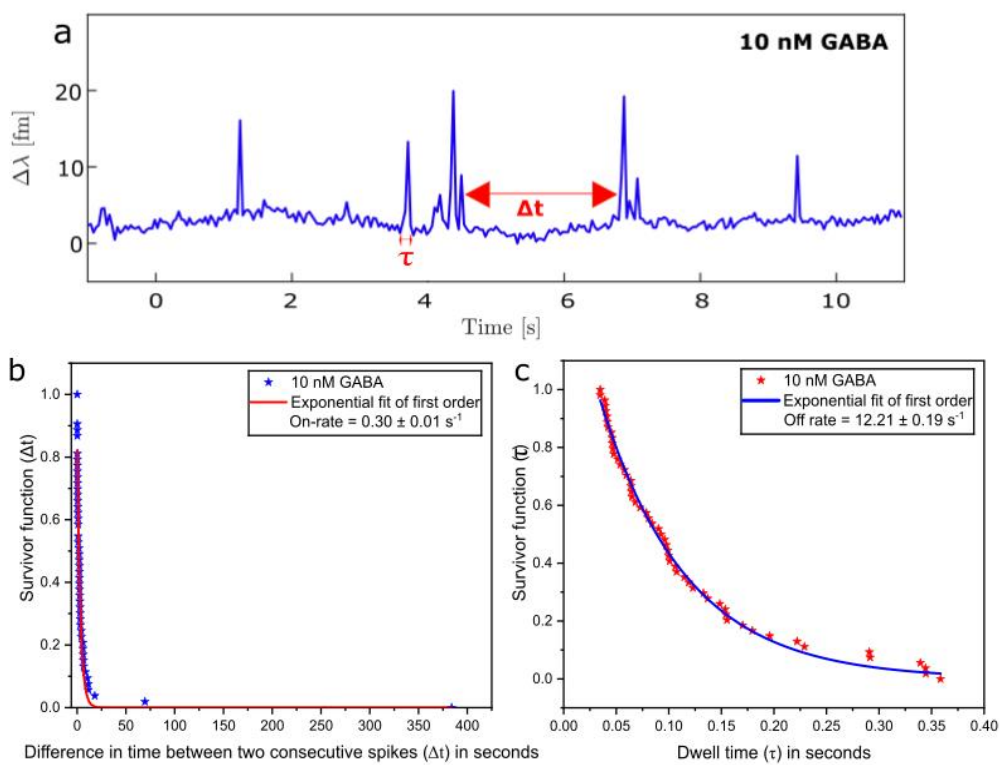

**Fig. S6 : Kinetics of GABA interaction with gold nanostars (AuNSs) a,** Spike signals are obtained when 10 nM GABA interact with AuNSs. **b,** Survivor function plots for the time separation of consecutive spikes and the corresponding exponential fits of first order. **c,** Survivor function plots for the duration of spikes (dwell time) for 10 nM GABA and its exponential fits.

The survivor function S($\Delta t$)=P(T>$\Delta t$)) represents the probability that no binding event occurs within the time interval $\Delta t$ between two consecutive spikes. By plotting this survivor function for the inter-event time distribution, the on-rate of neurotransmitter interaction with the nanoparticles can be extracted from the slope of the exponential fit. Similarly, replacing $\Delta t$ with the dwell time $\tau$ in the same formalism allows determination of the off-rate, which characterizes the duration of molecule-nanoparticle binding (see equation S1).

Fig. S6 a shows spike signals obtained from the interaction of 10 nM GABA with CTAB-coated gold nanostars (AuNSs). Here, $\Delta t$ denotes the time interval between consecutive events, and $\tau$ indicates the event dwell time. The corresponding survivor function plots for $\Delta t$ and $\tau$ are presented in supplementary Figs. S6b and S6c, respectively, both fitted with first-order exponential decay functions consistent with Poissonian kinetics. The on-rate derived from the time separation survivor function is 0.30 ± 0.01 s$^{-1}$, while the off-rate obtained from the dwell time survivor function is 12.21 ± 0.19 s$^{-1}$.

At GABA concentrations below 10 nM, fewer events were observed, likely due to the presence of only a single carboxylate group limiting interaction frequency. Consequently, statistical analysis and reliable rate extraction at these lower concentrations were not feasible.

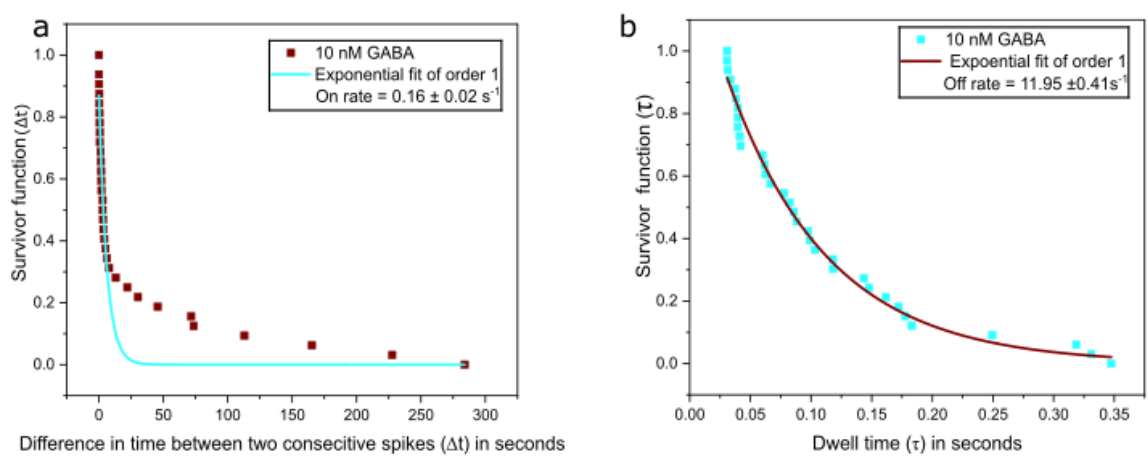

**Fig. S7: Kinetics of GABA interaction with gold nanorods (AuNRs).** *a,* Representative spike signals recorded during the interaction of 10 nM GABA with AuNRs. *b,* Survivor function plot of time intervals between consecutive binding events (Δt), with single-exponential fits used to extract the association on-rate. *c,* Survivor function plot of spike durations (dwell time, τ), with corresponding exponential fits to determine the dissociation off-rate.

The survivor function plots for GABA binding to CTAB-stabilized gold nanorods (AuNRs) at 10 nM are shown in Figst. S7 a and S7 b, corresponding to the time interval between consecutive events (Δt) and the event duration (dwell time, τ), respectively. The Δt distribution was fitted with a single-exponential decay function, yielding an association on- rate of $0.16 \pm 0.02$ s$^{-1}$. The dwell time distribution was similarly fitted, resulting in a dissociation off-rate of $11.95 \pm 0.41$ s$^{-1}$. For comparison, the on-rate for interaction of GABA to gold nanostars (AuNSs) under identical conditions was $0.30 \pm 0.01$ s$^{-1}$, indicating a greater number of binding events and a higher surface interaction probability on AuNSs. The slightly higher off-rate observed for AuNSs ($12.21 \pm 0.19$ s$^{-1}$) suggests longer event durations and greater hotspot saturation. These results highlight the enhanced sensitivity of AuNSs in the optoplasmonic WGM sensing configuration, as compared to AuNRs.

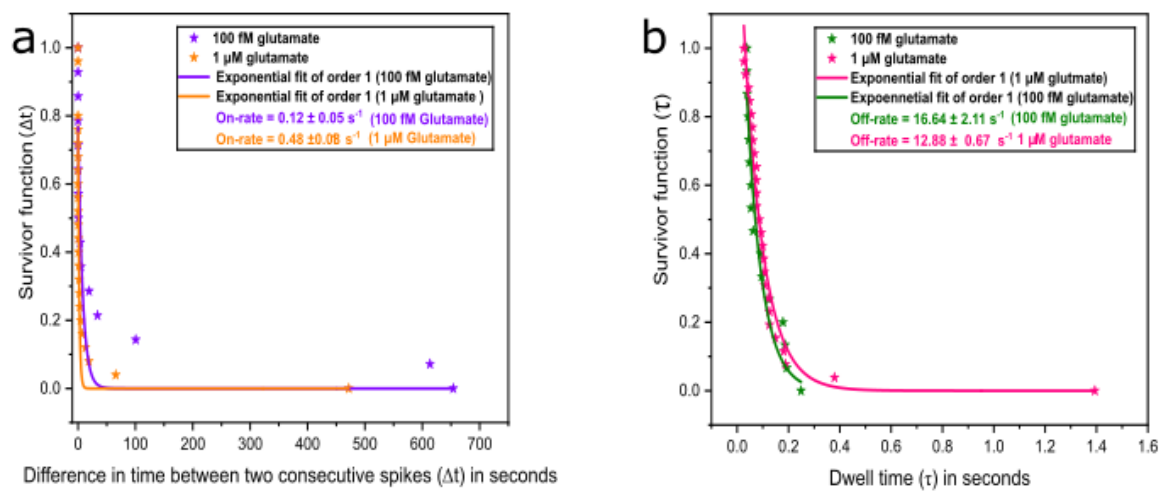

**Fig. S8: Concentration-dependent kinetics of glutamate interaction with gold nanostars (AuNSs).** *a,* Survivor function plots showing the time intervals (Δt) between consecutive binding events at glutamate concentrations of 100 fM and 1 μM. The data are fitted with single-exponential decay functions to extract the association on-rates. *b,* Survivor function plots of dwell times (τ) at the same concentrations, with corresponding exponential fits used to determine the dissociation off- rates. These measurements reveal the concentration dependence of glutamate binding kinetics at the single-molecule level.

Glutamate, which contains two carboxylate groups, exhibits more frequent transient interactions with carbonate anions chemisorbed on gold nanostars (AuNSs) than GABA, as indicated by a higher frequency of discrete spike signals. To quantify these interactions, kinetic measurements were performed at glutamate concentrations of 100 fM and 1 μM. Although the optoplasmonic WGM sensor demonstrated single-molecule sensitivity at 10 aM, the number of recorded events at this concentration was insufficient for statistical analysis. Therefore, kinetic fitting was performed on datasets acquired at 100 fM and 1 μM.

Survivor function analyses of time intervals between events (Δt) and event durations (dwell times, τ) were fitted with first-order exponential decay functions. The mean association on-rate at 100 fM was determined to be 0.12 ± 0.05 s$^{-1}$, increasing to 0.48 ± 0.08 s$^{-1}$ at 1 μM, consistent with a higher number of molecules interacting with the sensor surface at increased concentrations. The dissociation off-rate, decreased from 16.64 ± 2.11 s$^{-1}$ at 100 fM to 12.68 ± 0.67 s$^{-1}$ at 1 μM. This decrease in off-rate is attributed to the partial saturation of plasmonic hotspots on the AuNSs surface at higher concentrations, which increases the competition among molecules and stabilizes individual binding events. The extended dwell times observed at higher concentration thus reflect enhanced molecular occupancy at limited hotspot sites.

## 7. FDTD Simulations

The geometrical features of the plasmonic particle are tailored and optimized to have manifold enhancement in the spatial confinement of the electric field for enhanced nanoscale light-matter interaction with the analyte. It has been observed that the localized surface plasmon resonances (LSPRs) of noble metal nanoparticles, originated from the collective oscillations of conduction electrons, are strongly influenced by the size, shape, composition, and the surrounding environment. In our quest for the development of advanced single molecule optoplasmonic sensors we also explore the tailored metal nanoparticles such as nanorods and nanostars where the latter class have sharp-tipped spikes projecting from a core with engineered geometries befitting to the spectral region of interest in order to investigate the conformational dynamics, where huge enhancement of electric field localization could be attained in addition to the conventional plasmonic resonant effects as shown in Fig. S9 (a,b,c,d).

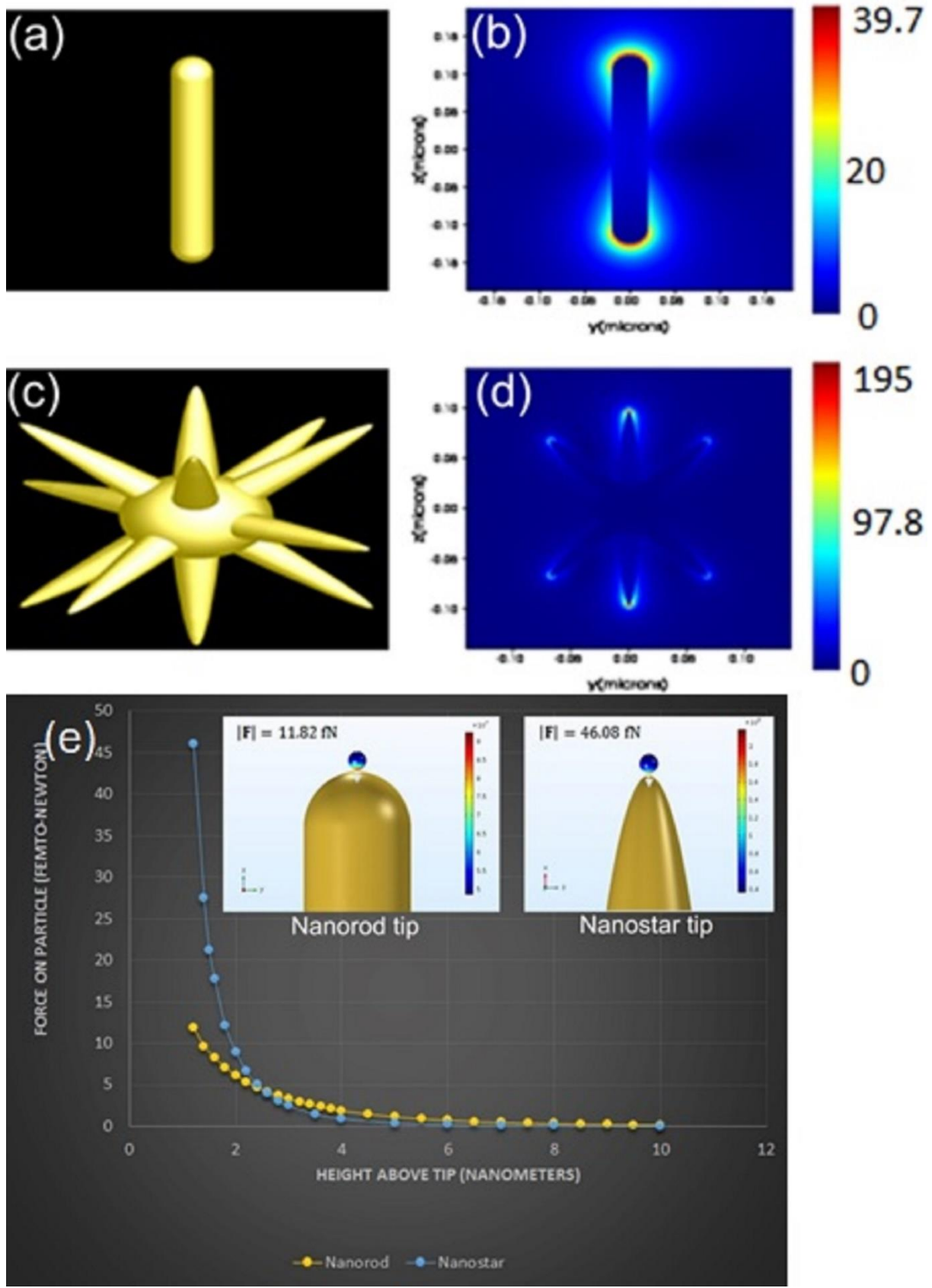

**Fig. S9: Comparison of electric field enhancement at a resonant peak of 800 nm**. (a) – (b) Plasmonic nanorods having rod height of 45 nm and radius 6 nm. (c)-(d). Plasmonic nanostars with sharp spikes having core diameter 60 nm and spike length 88 nm. (e) Comparison of the trapping force on a small dielectric particle with radius 1 nm for varying gap respectively from a plasmonic nanorod and nanostar. [Inset: Plasmonic nanorod (left) with height of 45 nm and radius 6 nm and Plasmonic nanostar (right) with core diameter 60 nm and spike length 88 nm.]

The sharp-tipped plasmonic structures have a field enhancement due to the dipolar fields of the particle plasmon resonance as well as an increase in the local field by the lighting-rod effect. The lightning-rod effect is a purely geometrical factor causing the dipolar fields of particles to be concentrated near their sharp tips. The lighting-rod effect also enables us to experimentally collect the signal detuned from the conventional plasmonic resonance itself which make these structures unique and highly attractive for this project. Since the large surface curvature of these spiked plasmonic particles possess multiple hotspots, the surface charges and local fields are concentrated and confined near the particle tips with sharp surface features. As the normal to surface component of the electric field near the surface of the sharp tips increases, there is an enhanced generation of quantum optically induced hot electrons with large energies. By means of the Maxwell stress tensor method we also investigated the total force on a dielectric particle (1 nm size) at the plasmon resonance for varying gap of plasmonic particle and the interacting dielectric particle (Fig. S9(e)). It is observed that the plasmonic nanostar traps the particles much more strongly within a 2.4 nm range. As shown in the plot, for an input power of 1E6 $W/cm^2$ the force could be realized in femto-Newtons with almost a 4-fold enhancement for the smaller gaps.

## 8. Computational details for DFT

Au22 cluster with 2 layers (the first layer has 14 Au atoms, and second layer has 8 Au atoms) was created by cleaving (111) surfaces from gold crystal. A third layer was judged to be unnecessary because of the local nature of the chemisorption phenomena and vibrational frequencies. We fixed the Au using the experimental lattice parameter of 4.078 Å. All geometric optimizations, frequency calculations, counterpoise (CP) corrections[1], basis set superposition error (BSSE) corrections and complexation energy computations were performed by using the Gaussian 16 series of programs[2]. All structures were optimized in the gas phase at the (U)M06-2X level of theory[3] in which all metal atoms were described with the lanl2dz basis set[4] and 6-31g** basis used for all non-metallic atoms. Lanl2dz basis set treats explicitly the outer $5s^2 5p^6 5d^{10} 6s^1$ electrons in gold atoms by a double-zeta basis set while the inner core electrons were replaced by the relativistic effective core potential, RECP, of Hay and Wadt[4]. For the geometry optimization and the corresponding frequency calculations, a tight convergence criterion and an ultrafine integration grid (99,590) were adopted.